\pdfoutput=1

\documentclass{PoS}
\usepackage{amsmath,amssymb}

\newcommand{\mgluino}{m_{\tilde{g}}}
\newcommand{\mstopone}{m_{\tilde{t}_1}}
\newcommand{\mstoptwo}{m_{\tilde{t}_2}}
\newcommand{\msquark}{m_{\tilde{q}}}

\title{Precision calculations for Higgs-boson production at hadron colliders}

\ShortTitle{Precision calculations for Higgs-boson production at hadron colliders}

\author{\speaker{Nikolai Zerf}$^{a,b}$\\
       \llap{$^a$}Institut f\"ur Theoretische Teilchenphysik, KIT\\
       \llap{$^b$}Department of Physics, University of Alberta\\
       \email{zerf@ualberta.ca}}

\abstract{The next-to-next-to-leading order (NNLO) calculation of the effective Higgs gluon coupling $C_1$ for the lightest Minimal Supersymmetric Standard Model (MSSM) Higgs boson $h^0$ is presented.
Selected numerical results for the total production cross section of $h^0$ via gluon fusion at the LHC within the effective field theory approach are shown in the NNLO approximation.}

\FullConference{``Loops and Legs in Quantum Field Theory'' 11th DESY Workshop on Elementary Particle Physics \\
                 April 15-20, 2012\\
                 Wernigerode, Germany}

\begin{document}

\section{Introduction}
The discovery of a new particle with a mass of about 125~GeV at the LHC~\cite{ATLAS-Higgs,CMS-Higgs} may be explained as discovery of the Standard Model (SM) Higgs boson.
But it is also possible that the lightest MSSM Higgs boson $h^0$ has been found.
The identification of the new particle as SM or MSSM Higgs boson requires precise theoretical predictions for its production and decay.
A lot of work has already gone into the more and more improved predictions of production cross sections and decay rates,
which goes hand in hand with including effects arising at higher orders in perturbation theory.
A short review of the current status concerning the production cross sections of the Higgs boson has been given in the talk connected to this proceedings contribution.
It was mainly based on the information given in the comprehensive reviews~\cite{Dittmaier:2011ti} and~\cite{Dittmaier:2012vm}.
Such a review will be skipped here in order to be able to focus on the new contributions presented at the conference, which have been published in~\cite{Pak:2010cu,Pak:2012xr}.

In detail this concerns the calculations of the total production cross section of $h^0$ via gluon fusion at next-to-next-to-leading order (NNLO) within the effective theory framework.
In more detail this calculation is enabled by the determination of the effective Higgs gluon coupling $C_1$ up to three loops
in the strong coupling constant $\alpha_s$ in the MSSM.

In the SM it is well known that radiative Quantum Chromodynamics (QCD) corrections to the loop induced gluon fusion process do play an important role.
The next-to-leading order (NLO) QCD corrections can increase the leading order (LO) cross section up to $+100$\%~\cite{Georgi:1977gs,Djouadi:1991tka,Dawson:1990zj,Spira:1995rr}
including a large uncertainty due to the variation of the cross section in dependence of the renormalization and factorization scale $\mu$.
The corrections appearing at NNLO further increases the cross section up to $+30\%$~\cite{Harlander:2002wh,Anastasiou:2002yz,Ravindran:2003um} compared to the NLO predictions.
They have first been calculated using the effective field theory approach, where the mass of the top quark is infinitely heavy.
It turned out that after including these NNLO corrections the scale variation is strongly reduced and
one can obtain a trustable prediction using perturbation theory.
By taking finite top mass effects at the NNLO level into account,
it has been checked,
that the used effective field theory (EFT) leads to reliable
predictions~\cite{Harlander:2009bw,Pak:2009bx,Harlander:2009mq,Pak:2009dg,Harlander:2010my,Pak:2011hs} as long as the Higgs mass stays below about 200~GeV.

The calculations of the production cross sections of $h^0$ in the MSSM at NLO has been accomplished by many groups using different methods including bottom induced contributions,
which can be large compared to the SM ones~\cite{Spira:1995rr} due to a $\tan{\beta}$ enhanced bottom Yukawa coupling~\cite{Harlander:2003bb,Harlander:2004tp,Anastasiou:2008rm,Degrassi:2008zj,Degrassi:2010eu,Harlander:2010wr,Muhlleitner:2010nm,Muhlleitner:2006wx,Bonciani:2007ex}.

Before the publication of Refs.~\cite{Pak:2010cu,Pak:2012xr} the NNLO corrections had to be approximated by the corresponding SM corrections~\cite{Harlander:2004tp,Harlander:2003kf},
because the three loop part of the effective Higgs coupling $C_1$ was not known in the MSSM.
In a first step this missing piece has been calculated with degenerated masses for supersymmetric particles (sparticles) in Ref.~\cite{Pak:2010cu}.
Ref.~\cite{Pak:2012xr} extends this calculation to apply to more general supersymmetric mass spectra.

The layout of this proceedings contribution will be as follows:
In chapter two we shortly review the calculation of the effective Higgs gluon coupling at three loop order.
In chapter three we show the connection between $C_1$ and the production cross section of $h^0$,
which will be evaluated numerically in chapter four in certain scenarios at NNLO.
The last chapter contains summary and conclusions.

\section{Effective Higgs gluon coupling}
\begin{figure}[htbp]
 \centering
\begin{tabular}{ccc}
 \raisebox{-1.2cm}{\includegraphics[scale=0.35]{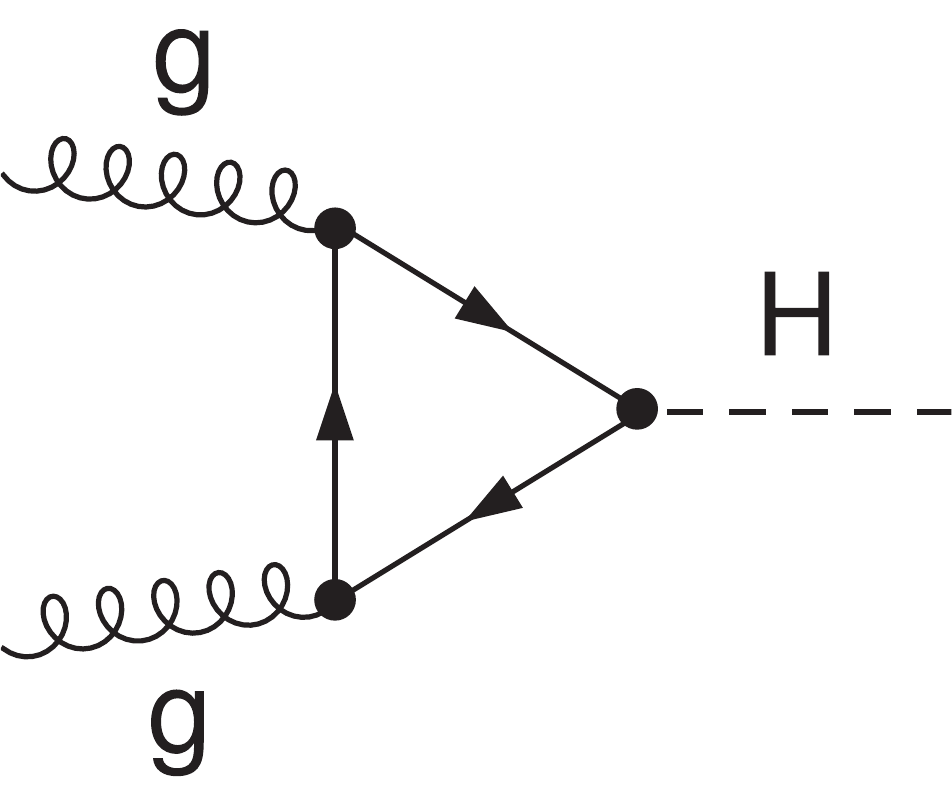}} & \raisebox{-1cm}{\includegraphics[scale=0.35]{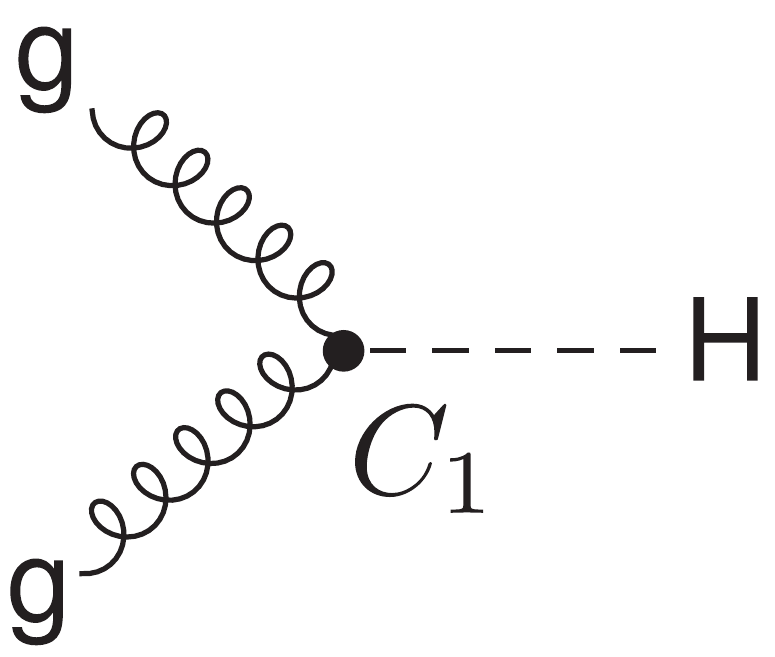}} &
\begin{tabular}{|c|c|c|c|}\hline
\multicolumn{2}{|c|}{particles}&\multicolumn{2}{|c|}{sparticles}\\\hline
$g$&gluon&$\tilde{g}$&gluino\\\hline
$t$&top quark& $\tilde{t}_i$&top squark\\\hline
$q$&light quarks& $\tilde{q}_i$&light squarks\\\hline\hline
\multicolumn{2}{|c|}{auxiliary particles}&\multicolumn{2}{|c}{}\\\cline{1-2}
$c$&ghost& \multicolumn{2}{|c}{}\\\cline{1-2}
$\epsilon$&epsilon scalar& \multicolumn{2}{|c}{} \\\cline{1-2}
\end{tabular}\\
 (a) & (b) & (c)
\end{tabular}
\caption{(a-b) LO Feynman diagrams for the process $g g\rightarrow H$ in the SM.
(a) In the full theory the Higgs boson couples via a top quark loop to the Higgs boson.
(b) In the EFT a direct coupling of gluons and the Higgs boson emerges.
(c) Particles appearing in the loops during the calculation of $C_1$ in the MSSM.}
\label{Figure1}
\end{figure}

\begin{figure}[htbp]
\centering
 \includegraphics[scale=0.80]{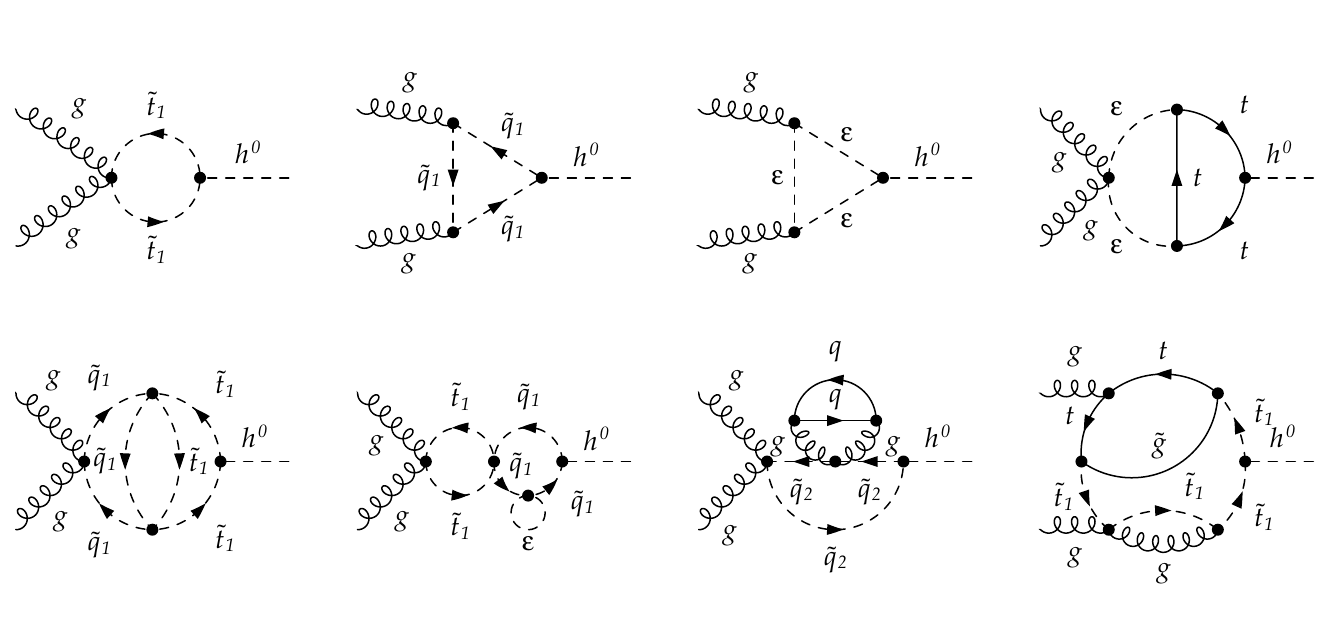}
\caption{Feynman diagrams contributing to $C_1$ in the MSSM.}
\label{Figure2}
\end{figure}

The effective Higgs gluon coupling arises when one integrates out all particles heavier than the Higgs boson $H/h^0$ (SM/MSSM) from the SM/MSSM Lagrange density.
Thus in the SM one has to integrate out the top quark only.
In the MSSM in addition all supersymmetric particles have to be integrated out.
The only relevant remnant of those particles in the Lagrange density is the first term on the r.h.s. of the following equation:
\begin{eqnarray}
  {\cal L}_{Y,\rm eff} &=& -\frac{h^0}{v^0} C_1^0 {\cal O}_1^0
  + {\cal L}_{QCD}^{(5)}
  \,,
  \label{eq::leff}
\end{eqnarray}
where the Operator ${\cal O}_1^0$ is built up by the gluonic field strength tensor $G^{0,\mu\nu}$ as follows:
\begin{eqnarray}
  {\cal O}_1^0 &=& \frac{1}{4} G_{\mu\nu}^0 G^{0,\mu\nu}\,,
\end{eqnarray}
and ${\cal L}_{QCD}^{(5)}$ is the well-known five flavour QCD Lagrange density.
To ensure the agreement of the predictions of the effective Lagrange density ${\cal L}_{Y,\rm eff}$ and the
 the supersymmetric QCD (SQCD) Lagrange density in the limit where the mass $M_{h}$ of the Higgs boson is much smaller than the mass of the top quark and all supersymmetric particles,
one has to determine the effective coupling $C_1^0$ by comparing equivalent Green's functions of both Lagrange densities incorporating a single Higgs boson and gluons.
As simplest Green's function one can consider the transition of two gluons to the Higgs boson like depicted for the SM case in Fig.~\ref{Figure1}(a-b).
The comparison of the two Green's functions is performed in the on-shell (OS) production kinematics.
That means $(p_1+p_2)^2=M_{H}^2$ holds for the four momenta $p_1$ and
$p_2$ of the two gluons with $p_1^2=p_2^2=0$.
In order to determine $C_1$ in the SM at LO one has to calculate the Feynman diagram in Fig.~\ref{Figure1}(a)
(and the one with reversed fermion lines) in the Limit $M_{H} \ll m_t$, where $m_t$ is the top quark mass.
$C_1$ is directly given as the first term in the corresponding expansion for a small Higgs masses and is proportional to $\alpha_s$ at LO.
For a more detailed description how to extract $C_1$ the reader may be referred to Ref.~\cite{Chetyrkin:1997un}.
In the MSSM case additional diagrams including sparticles have to be considered.
A list of all particles appearing inside loops can be found in Fig.~\ref{Figure1}(c).
For a determination of $C_1$ at NLO/NNLO all relevant two/three loop diagrams contributing to this process have to be calculated.
Example diagrams for the MSSM case are shown in Fig.~\ref{Figure2}.
Due to the expansion in the hierarchy $M_{H} \ll m_t$ the problem is reduced to the calculation of two/three loop tadpole integrals with more than one scale.
At two loops the analytic results of the integrals for arbitrary masses are known~\cite{Davydychev:1992mt} and $C_1$ can be calculated exactly to this order.
At three loops this is not the case and one has to reduce the number of scales in one diagram to one.
This can be achieved assuming nearly degenerated and/or very different (hierarchic) masses.
In the first case one can Taylor expand in small mass differences in the second case one can apply an asymptotic expansion.
Combinations of both methods are possible, too.
Once the number of scales has been reduced to one,
the appearing integrals can be solved completely automatically with the {\tt FORM} program {\tt MATAD}~\cite{hep-ph/0009029}.
Because of the large number of diagrams, which have to be calculated, an automatic setup is mandatory.
For the generation of Feynman diagrams we have used {\tt QGRAF}~\cite{Nogueira:2006pq}, followed by {\tt q2e} and {\tt exp}~\cite{Seidensticker:1999bb},
which provide an automatic application of asymptotic expansions.

Besides the problem of multiple scales in MSSM diagrams,
additional subtleties arise, when one uses Dimensional Reduction (DRED) as regulator,
in order not to spoil supersymmetry (SUSY):

Because the EFT cross section~\cite{Harlander:2000mg,Harlander:2002wh,Kilgore:2002yw} was calculated using Dimensional Regularization (DREG),
one has to ensure, that the epsilon scalars (which are the $2\epsilon$-dimensional components of gluon in DRED) are integrated out,
so that the EFT remains regulated in DREG.
This can be achieved by giving the epsilon scalars a mass which is formally much larger than the Higgs boson mass.
Due to the fact that SUSY is softly broken in the MSSM, the epsilon scalars do acquire a non vanishing mass via radiative corrections anyway.
This mass has to be understood as evanescent parameter in the Lagrange density.

In addition to that an evanescent coupling $\Lambda_{\epsilon}^2$ of the epsilon scalars to the Higgs boson does appear via radiative corrections.
This coupling has to be properly renormalized\footnote{for more details the reader shall be referred to Refs.~\cite{Pak:2010cu,Pak:2012xr}}
at the two-loop level when one considers $C_1$ at three loops.

Apart from these subtleties one has to renormalize all particle masses for the three-loop calculation of $C_1$ at the two-loop level,
which are well known in the $\overline{\text{DR}}$ scheme~\cite{Hermann:2011ha}.
Except for the gluino and epsilon scalar mass,
where only the one loop renormalization constants are required.

To arrive at a finite result for $C_1$ one requires the $\overline{\text{MS}}$ renormalization constant of the effective Operator $Z_{11}$,
the decoupling constant of the gluon field $\zeta_3^0$ and the decoupling constants of $\alpha_s$ $\zeta_{\alpha_s}$ at the two loop level.

Up to now $C_1$ has been calculated at the NNLO assuming the following hierarchies:
\begin{eqnarray}
  ({\rm h1}) && \msquark \approx \mstopone \approx \mstoptwo \approx \mgluino
  \gg m_t
  \,,\nonumber\\
  ({\rm h2}) && \msquark \approx \mstoptwo \approx \mgluino \gg \mstopone \gg
  m_t
  \,,\nonumber\\
  ({\rm h3}) && \msquark \approx \mstoptwo \approx \mgluino \gg \mstopone
  \approx m_t
  \,,
  \label{eq::hier}
\end{eqnarray}
Where the mass of the epsilon scalar was treated like being much smaller than the masses appearing above.
The renormalization was done in the OS scheme,
so that in the final expression of $C_1$ the limit $M_{\epsilon}\rightarrow0$ can be taken.

To ensure that the described procedure consistently leads to $C_1$,
it has been applied in the SM case.
That means instead of using the common DREG regulator, DRED has been used like written above.
Although the application of DRED to non supersymmetric theories leads to additional evanescent couplings,
the well known SM (QCD) result was obtained when $C_1$ was expressed in terms of the five flavour QCD coupling $\alpha_s^{(5)}$~\cite{Pak:2010cu}.

Another check for the calculation of $C_1$ was enabled by the three loop calculation of $\zeta_{\alpha_s}$ and its connection to $C_1$ via the Low Energy Theorem (LET)~\cite{Kurz:2012ff}.
It turned out that with the renormalized version of the LET one is able to calculate $C_1$ from $\zeta_{\alpha_s}$
without even knowing the renormalization constant $\delta \Lambda_{\epsilon}^2$ of the evanescent coupling of the epsilon scalars to the Higgs boson.
However, both ways lead, of course, to the same result for $C_1$.

Explicit results for $C_1$ can be found in Ref.~\cite{Pak:2012xr} and Ref.~\cite{progdata}.

\section{Calculating Cross Sections}
The calculation of the production cross section of a light Higgs boson using the effective Lagrangian ${\cal L}_{Y,\rm eff}$ could be done~\cite{Harlander:2002wh}
completely independent of the calculation of $C_1$, because the effective coupling factorizes in the formula for the total cross section:
\begin{eqnarray}
 \sigma_t(\mu)=\sigma_0\Big(\frac{3\pi}{c_1^{(0)}}\Big)^2C_1^2(\mu)\Sigma^{\prime}(\mu)\,.\label{XSection1}
\end{eqnarray}
Where
$C_1=-\frac{\alpha_s^{(5)}}{3\pi}\sum_{n=0}^{\infty}\Big(\frac{\alpha_s^{(5)}}{\pi}\Big)^nc_1^{(n)}$
and $\Sigma^{\prime}(\mu)$ contains the sum over all channels which in
turn are obtained by convolutions of the partonic cross sections with the particle distribution functions (PDFs).
Since $\sigma_0$ contains the full Higgs mass dependence\footnote{For the explicit formula in the MSSM see for e.g. Eq. (29) in Ref. \cite{Pak:2012xr}.} at LO,
$\sigma_0\alpha_s^{(5),2}\Big[\Sigma^{\prime}(\mu)\Big]_{\text{LO}}$ gives the exact LO cross section.
Thus in Eq.~(\ref{XSection1}) higher order corrections obtained in the EFT are added relative to the factorized,
exact leading order cross section, which improves the EFT prediction.

A further improvement of this prediction can be achieved via a separation of hard $\mu_h$ and soft scale $\mu_s$:
It is known that only the combination $C_1(\mu){\cal O}_1(\mu)$ is independent of the scale $\mu$ (up to neglected higher order corrections).
Introducing the factor\footnote{The explicit definition is given in Eq. (32) Ref. \cite{Pak:2012xr}.} $B^{(5)}$,
which depends on the coefficient of the beta function and $\alpha_s^{(5)}$ enables the definition of a separately scale invariant matching coefficient $C_g$ and
effective operator ${\cal O}_g$, which do only depend on $\mu$ due to neglected higher order corrections.
Thus one can choose a hard scale $\mu_h$ in the matching coefficient and a soft scale $\mu_s$ in the matrix elements of the effective Operator in order to avoid large logarithms.
In that case the cross section formula becomes:
\begin{eqnarray}
 \sigma_t(\mu_s,\mu_h)=\sigma_0C_g^2(\mu_h)\Sigma(\mu_s)\,.\label{XSection2}
\end{eqnarray}
Where $C_g = -\frac{3\pi}{ c_1^{(0)} } \frac{1}{B^{(5)}} C_1$ and $\Sigma=\big(B^{(5)}\big)^2\Sigma^{\prime}$.
Note that in the limit $\mu_h=\mu_s$ the $B$-Factor cancels and one obtains Eq.~(\ref{XSection1}).
Eq.~(\ref{XSection2}) holds for SM and MSSM predictions, as long as one uses the corresponding expressions for $\sigma_0$ and $C_g$.
However there is only one expression for $\Sigma(\mu_s)$.
Thus, as soon as the matching coefficient in the MSSM $C_1^{\text{SQCD}}$ is known, it is straightforward to determine the cross section $\sigma_t^{\text{SQCD}}(\mu_s,\mu_h)$ in the MSSM using $\Sigma(\mu_s)$ extracted from the SM cross section $\sigma^{\text{QCD}}_t(\mu_s,\mu_h)$.
The predictions in the next section are based on the predictions of the SM cross section $\sigma^{\text{QCD}}_t(\mu_s,\mu_h)$ obtained by the program {\tt Xsection},
which was used to produce the SM predictions of Ref.~\cite{Pak:2011hs}.

\section{Numerical Results}
\begin{figure}[t]
  \centering
  \begin{tabular}{cc}
    \includegraphics[width=.48\linewidth]{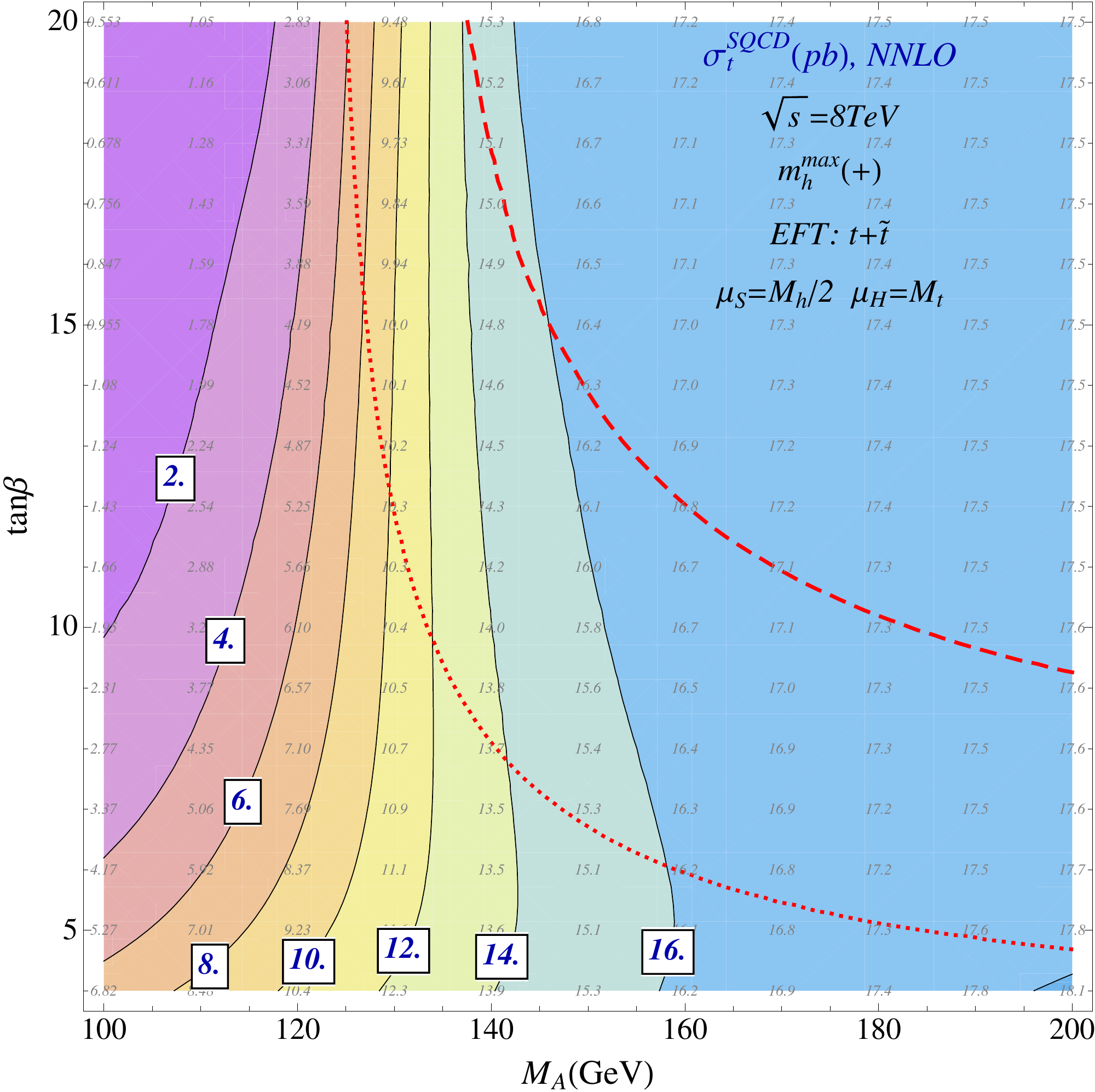}
    &
    \includegraphics[width=.48\linewidth]{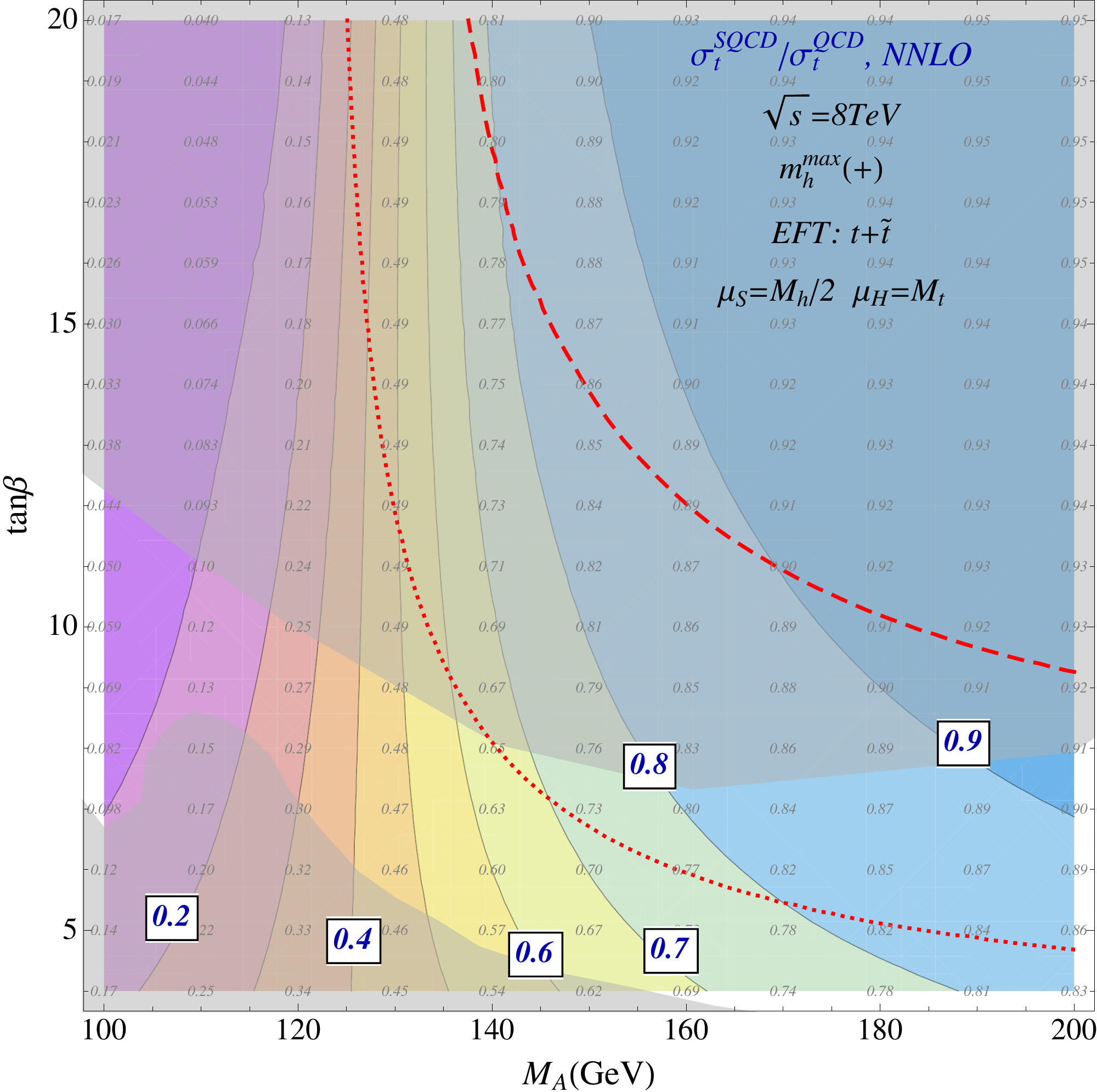}
    \\
    (a) & (b)
  \end{tabular}
  \caption[]{\label{Figure3}(a) $\sigma_t^{\rm SQCD}(\mu_s,\mu_h)$
    and (b) $\sigma^{\rm SQCD}_t/\sigma^{\rm QCD}_t$ as a function of $M_A$ and $\tan\beta$ in the
    $m_h^{\rm max}$ scenario.}
\end{figure}
In Fig.~\ref{Figure3}(a) the NNLO EFT cross section $\sigma^{\text{SQCD}}_t(\mu_s,\mu_h)$ for a hadronic center of mass energy of 8~TeV at the LHC is shown in pico barn assuming the $m_h^{\text{max}}$ scenario~\cite{Carena:2002qg}.
For the calculation the MSTW2008 PDFs set~\cite{Martin:2009iq} and the result for $C_1^{\text{SQCD}}$ obtained in the hierarchy (h1) has been used.
The Higgs boson mass $M_{h}$ between the red dotted (lower) line  and red dashed (upper) line varies between 123 and 129~GeV.
The strong decrease of the cross section for small values of $M_A$ and large values of $\tan\beta$ results mainly from a decrease of the top quark Higgs boson coupling due to the factor $\propto\frac{\cos\alpha}{\sin\beta}$,
which is also present in the corresponding top squark Higgs coupling.

The ratio of the SM EFT cross section and the MSSM one is shown in Fig.~\ref{Figure3}(b).
Here the SM cross section depends only via the mass of the Higgs boson on the parameters $M_A$ and $\tan\beta$.
That means for each point both cross section have been calculated using the mass of the lightest MSSM Higgs boson at this point.
The gray shaded areas are already excluded by {\tt LEP}~\cite{Schael:2006cr} and {\tt CMS}~\cite{Chatrchyan:2012vp}.
Only the parameter sets corresponding to the unshaded area between the two red curves are in agreement with current experiments.
From the figure one can clearly see, that for growing values of $M_A$ the cross section becomes more and more SM like.
For small values of $\tan\beta$ and $M_A$ $\sigma_t^{\text{SQCD}}(\mu_s,\mu_h)$ can be much smaller than $\sigma_t^{\text{QCD}}(\mu_s,\mu_h)$.

\begin{figure}[t]
  \centering
  \begin{tabular}{cc}
    \includegraphics[width=.48\linewidth]{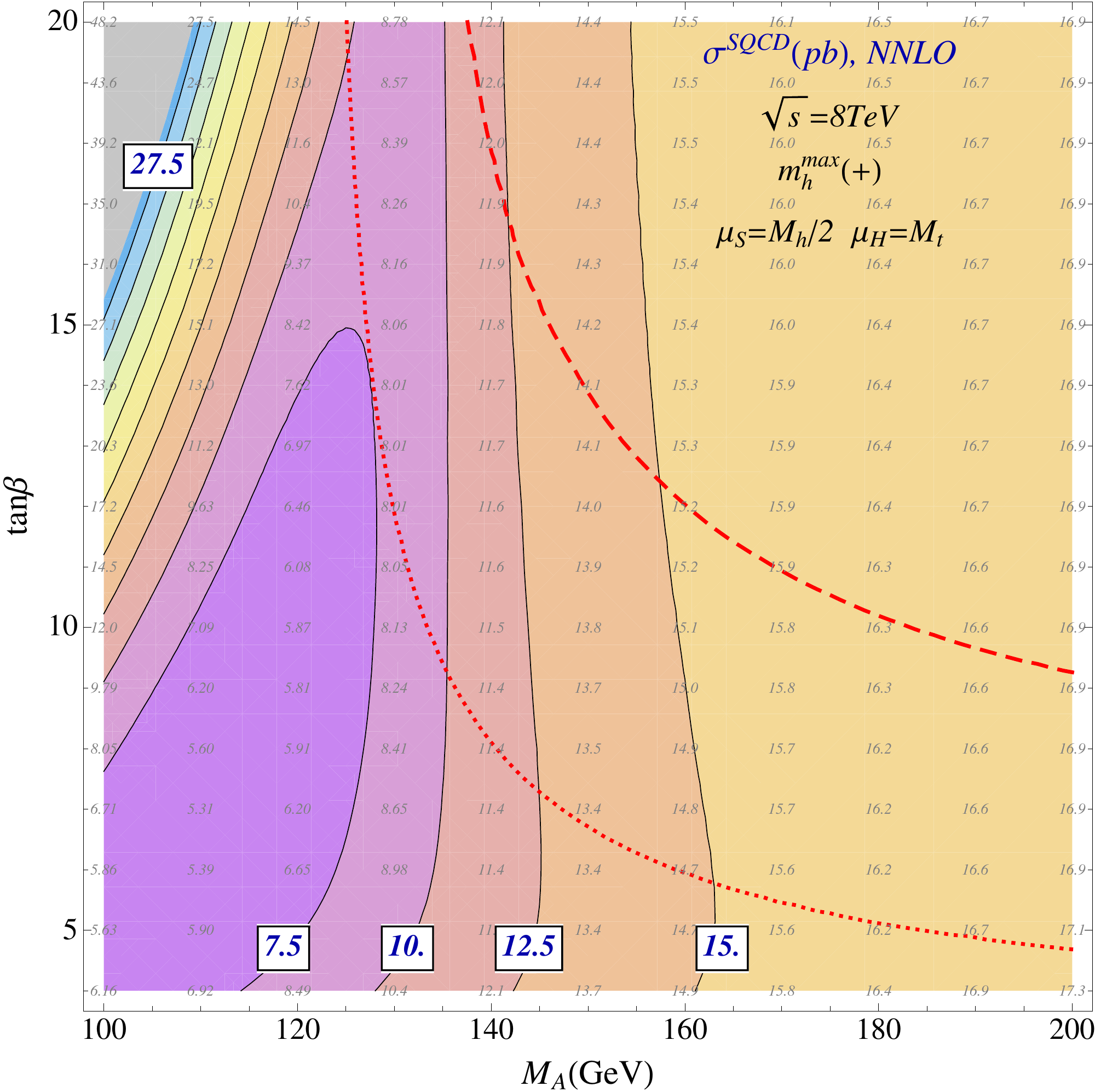}
    &
    \includegraphics[width=.48\linewidth]{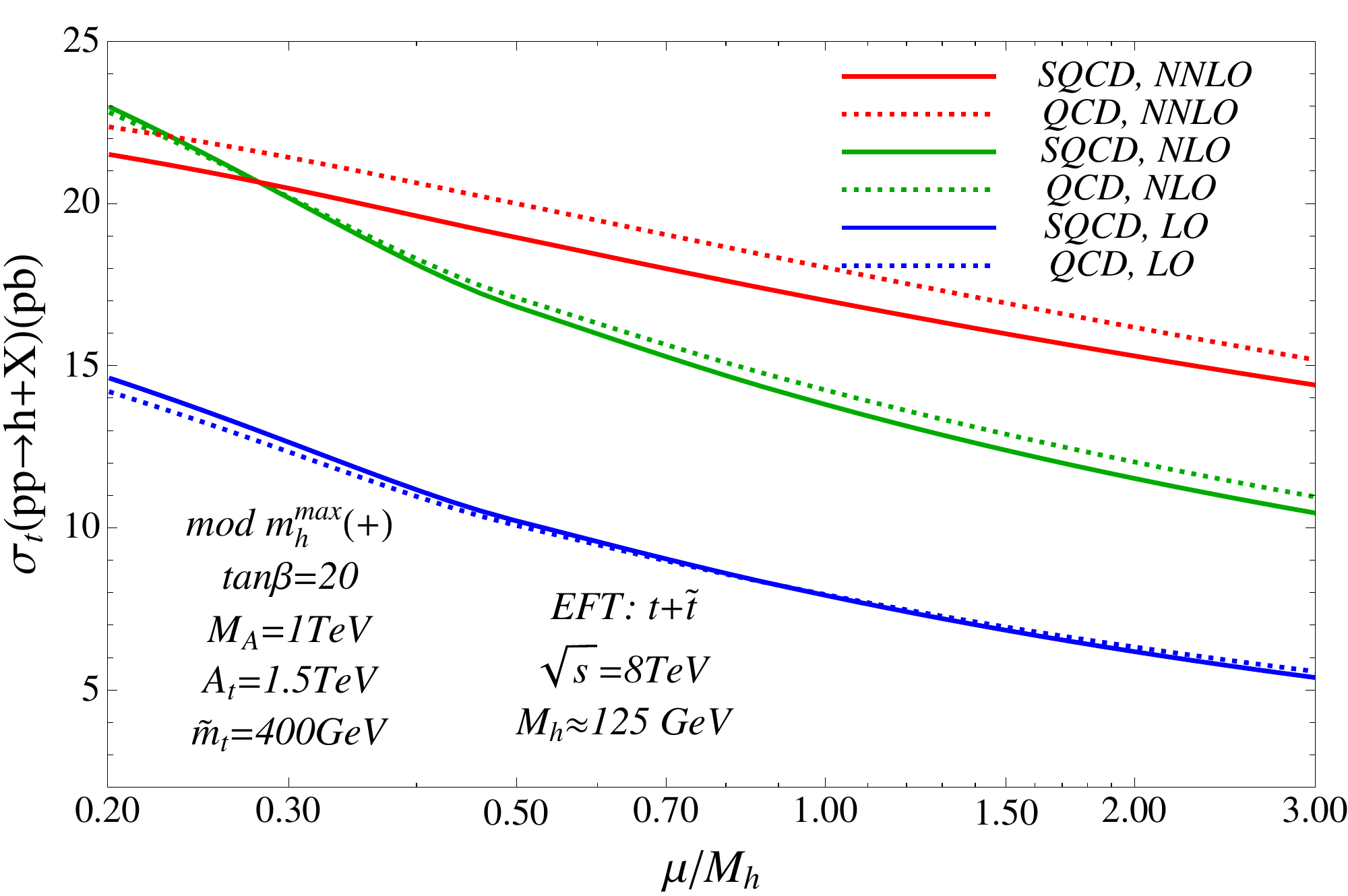}%
    \\
    (a) & (b)
  \end{tabular}
  \caption[]{\label{Figure4}(a) $\sigma^{\rm SQCD}$
    including electroweak and finite bottom mass induced contribution at NLO. (b) Behaviour of $\sigma^{\rm SQCD}_t$ under scale variation in the
    modified $m_h^{\rm max}$ scenario.}
\end{figure}

At this point it is very important to note that in the used EFT approach no finite bottom mass effects have been considered.
That means the bottom quark is treated as massless particle and does not couple to the Higgs boson.
However for large $\tan\beta$ values (and small $M_A$ values) it is know that the bottom Yukawa coupling to the Higgs boson can be enhanced.
Thus one can expect large bottom quark induced contributions in this parameter region.
In fact this is what other groups confirmed at NLO~\cite{Degrassi:2010eu,Harlander:2010wr,Anastasiou:2008rm} and in order not to miss those contributions they have been included in Fig.~\ref{Figure4}(a).
In addition the contributions of the known electroweak NLO corrections have been included.
More details about how to combine $\sigma^{\rm SQCD}_t$ with the mentioned contributions to obtain $\sigma^{\rm SQCD}$ can be found in Ref.~\cite{Pak:2012xr}.

We want to note that the usual estimation of the NNLO part of $C_1^{\text{SQCD}}$ by the one of $C_1^{\text{QCD}}$ results in a good approximation of the cross section in the $m_h^{\text{max}}$ scenario.
Where the error is clearly below the one percent level.

In Fig.~\ref{Figure4} $\sigma_t(\mu)$ is shown in dependence of the common scale $\mu=\mu_s=\mu_h$ in the modified $m_h^{\text{max}}$ scenario,
which was introduced in Ref.~\cite{Pak:2012xr}.
The dotted/solid curves represent the SM/MSSM result,
where from bottom to top the blue/green/red lines show the LO/NLO/NNLO predictions.
One can see a reduction of scale dependence
when one increases the order of perturbation theory.
The observation that at NNLO there is a clearly visible difference between the QCD and SQCD curve makes clear,
that the above statement concerning the estimation of the NNLO part of $C_1^{\rm SQCD}$ given for $m_h^{\text{max}}$ scenario does not hold for this scenario.
In fact the scenario was constructed in such a way,
that besides providing a Higgs mass in the current LHC range,
we have a light top squark mass of about 370~GeV,
whereas all other sparticles have masses around 1~TeV.
This enables diagrams including the light top squark $\tilde{t}_1$ to give sizable contributions to $C_1^{\rm SQCD}$,
 leading to a strong deviation from the NNLO contribution in the SM.

\section{Summary}
In this proceedings contribution a short introduction concerning the Higgs boson production via gluon fusion was given.
Furthermore the calculation of the NNLO contributions to the effective Higgs gluon coupling $C_1^{\rm SQCD}$ in the MSSM was presented in a compact form.
Selected numerical results implementing this contributions were shown.

In the $m_h^{\text{max}}$ scenario the SM NNLO part of $C_1^{\text{QCD}}$ can be used to estimate the NNLO part of $C_1^{\text{SQCD}}$.
However, it has been shown that in the presents of light sparticles like light top squarks,
this estimation can fail.

\acknowledgments
I want to thank Alexey Pak for his great work on {\tt Xsection} and Matthias Steinhauser for the steady support throughout my Ph.D. time in Karlsruhe.


\begin{thebibliography}{99}
\bibitem{ATLAS-Higgs}
   ATLAS Collaboration, arXiv:1207.7214.

\bibitem{CMS-Higgs}
   CMS Collaboration,  arXiv:1207.7235.

\bibitem{Dittmaier:2011ti}
  S.~Dittmaier, C.~Mariotti, G.~Passarino, R.~Tanaka {\it et al.},
   [LHC Higgs Cross Section Working Group Collaboration],
  arXiv:1101.0593 [hep-ph].

\bibitem{Dittmaier:2012vm}
  S.~Dittmaier, C.~Mariotti, G.~Passarino, R.~Tanaka {\it et al.},
   [LHC Higgs Cross Section Working Group Collaboration],
  arXiv:1201.3084 [hep-ph].

\bibitem{Pak:2010cu}
  A.~Pak, M.~Steinhauser and N.~Zerf,
  Eur.\ Phys.\ J.\  C {\bf 71} (2011) 1602
  [arXiv:1012.0639 [hep-ph]].

\bibitem{Pak:2012xr}
  A.~Pak, M.~Steinhauser and N.~Zerf,
  arXiv:1208.1588 [hep-ph].

\bibitem{Georgi:1977gs}
  H.~M.~Georgi, S.~L.~Glashow, M.~E.~Machacek and D.~V.~Nanopoulos,
  Phys.\ Rev.\ Lett.\  {\bf 40} (1978) 692.

\bibitem{Djouadi:1991tka}
  A.~Djouadi, M.~Spira and P.~M.~Zerwas,
  Phys.\ Lett.\  B {\bf 264} (1991) 440.

\bibitem{Dawson:1990zj}
  S.~Dawson,
  Nucl.\ Phys.\  B {\bf 359} (1991) 283.

\bibitem{Spira:1995rr}
  M.~Spira, A.~Djouadi, D.~Graudenz and P.~M.~Zerwas,
  Nucl.\ Phys.\  B {\bf 453} (1995) 17,
  arXiv:hep-ph/9504378.

\bibitem{Harlander:2002wh}
  R.~V.~Harlander and W.~B.~Kilgore,
  Phys.\ Rev.\ Lett.\  {\bf 88} (2002) 201801,
  arXiv:hep-ph/0201206.

\bibitem{Anastasiou:2002yz}
  C.~Anastasiou and K.~Melnikov,
  Nucl.\ Phys.\  B {\bf 646} (2002) 220,
  arXiv:hep-ph/0207004.

\bibitem{Ravindran:2003um}
  V.~Ravindran, J.~Smith and W.~L.~van Neerven,
  Nucl.\ Phys.\  B {\bf 665} (2003) 325,
  arXiv:hep-ph/0302135.

\bibitem{Harlander:2009bw}
  R.~V.~Harlander and K.~J.~Ozeren,
  Phys.\ Lett.\  B {\bf 679} (2009) 467
  [arXiv:0907.2997 [hep-ph]].

\bibitem{Pak:2009bx}
  A.~Pak, M.~Rogal and M.~Steinhauser,
  Phys.\ Lett.\  B {\bf 679} (2009) 473
  [arXiv:0907.2998 [hep-ph]].

\bibitem{Harlander:2009mq}
  R.~V.~Harlander and K.~J.~Ozeren,
  JHEP {\bf 0911} (2009) 088
  [arXiv:0909.3420 [hep-ph]].

\bibitem{Pak:2009dg}
  A.~Pak, M.~Rogal and M.~Steinhauser,
  JHEP {\bf 1002} (2010) 025
  [arXiv:0911.4662 [hep-ph]].

\bibitem{Harlander:2010my}
  R.~V.~Harlander, H.~Mantler, S.~Marzani and K.~J.~Ozeren,
  arXiv:0912.2104 [hep-ph].

\bibitem{Pak:2011hs}
  A.~Pak, M.~Rogal and M.~Steinhauser,
  JHEP {\bf 1109} (2011) 088
  [arXiv:1107.3391 [hep-ph]].

\bibitem{Harlander:2003bb}
  R.~V.~Harlander and M.~Steinhauser,
  Phys.\ Lett.\  B {\bf 574} (2003) 258
  [arXiv:hep-ph/0307346].

\bibitem{Harlander:2004tp}
  R.~V.~Harlander and M.~Steinhauser,
  JHEP {\bf 0409} (2004) 066
  [arXiv:hep-ph/0409010].

\bibitem{Anastasiou:2008rm}
  C.~Anastasiou, S.~Beerli and A.~Daleo,
  Phys.\ Rev.\ Lett.\  {\bf 100} (2008) 241806
  [arXiv:0803.3065 [hep-ph]].

\bibitem{Degrassi:2008zj}
  G.~Degrassi and P.~Slavich,
  Nucl.\ Phys.\ B {\bf 805} (2008) 267
  [arXiv:0806.1495 [hep-ph]].

\bibitem{Degrassi:2010eu}
  G.~Degrassi and P.~Slavich,
  JHEP {\bf 1011} (2010) 044
  [arXiv:1007.3465 [hep-ph]].

\bibitem{Harlander:2010wr}
  R.~V.~Harlander, F.~Hofmann and H.~Mantler,
  JHEP {\bf 1102} (2011) 055
  [arXiv:1012.3361 [hep-ph]].

\bibitem{Muhlleitner:2010nm}
  M.~Muhlleitner, H.~Rzehak and M.~Spira,
  arXiv:1001.3214 [hep-ph].

\bibitem{Muhlleitner:2006wx}
  M.~Muhlleitner and M.~Spira,
  Nucl.\ Phys.\  B {\bf 790} (2008) 1
  [arXiv:hep-ph/0612254].

\bibitem{Bonciani:2007ex}
  R.~Bonciani, G.~Degrassi and A.~Vicini,
  JHEP {\bf 0711} (2007) 095
  [arXiv:0709.4227 [hep-ph]].

\bibitem{Harlander:2003kf}
  R.~Harlander and M.~Steinhauser,
  Phys.\ Rev.\  D {\bf 68} (2003) 111701
  [arXiv:hep-ph/0308210].

\bibitem{Chetyrkin:1997un}
  K.~G.~Chetyrkin, B.~A.~Kniehl and M.~Steinhauser,
  Nucl.\ Phys.\  B {\bf 510} (1998) 61,
  arXiv:hep-ph/9708255.

\bibitem{Davydychev:1992mt}
  A.~I.~Davydychev and J.~B.~Tausk,
  Nucl.\ Phys.\  B {\bf 397} (1993) 123.

\bibitem{hep-ph/0009029}
  M.~Steinhauser,
  Comput.\ Phys.\ Commun.\ \ {\bf 134} (2001) 335
  [hep-ph/0009029].

\bibitem{Nogueira:2006pq}
  P.~Nogueira,
  Nucl.\ Instrum.\ Meth.\  A {\bf 559}, 220 (2006).

\bibitem{Seidensticker:1999bb}
  T.~Seidensticker,
  arXiv:hep-ph/9905298.

\bibitem{Harlander:2000mg}
  R.~V.~Harlander,
  Phys.\ Lett.\  B {\bf 492} (2000) 74
  [arXiv:hep-ph/0007289].

\bibitem{Kilgore:2002yw}
  W.~B.~Kilgore and R.~V.~Harlander,
  arXiv:hep-ph/0205152.

\bibitem{Hermann:2011ha}
  T.~Hermann, L.~Mihaila and M.~Steinhauser,
  Phys.\ Lett.\ B {\bf 703} (2011) 51
  [arXiv:1106.1060 [hep-ph]].

\bibitem{Kurz:2012ff}
  A.~Kurz, M.~Steinhauser and N.~Zerf,
  arXiv:1206.6675 [hep-ph].

\bibitem{progdata}
{\tt http://www-ttp.particle.uni-karlsruhe.de/Progdata/ttp12/ttp12-26/}

\bibitem{Carena:2002qg}
  M.~S.~Carena, S.~Heinemeyer, C.~E.~M.~Wagner and G.~Weiglein,
  Eur.\ Phys.\ J.\ C {\bf 26} (2003) 601
  [hep-ph/0202167].

\bibitem{Martin:2009iq}
  A.~D.~Martin, W.~J.~Stirling, R.~S.~Thorne and G.~Watt,
  Eur.\ Phys.\ J.\  C {\bf 63} (2009) 189
  [arXiv:0901.0002 [hep-ph]].

\bibitem{Schael:2006cr}
  S.~Schael {\it et al.}  [ALEPH and DELPHI and L3 and OPAL and LEP Working
  Group for Higgs Boson Searches Collaborations],
  Eur.\ Phys.\ J.\ C {\bf 47} (2006) 547
  [hep-ex/0602042].

\bibitem{Chatrchyan:2012vp}
  S.~Chatrchyan {\it et al.}  [CMS Collaboration],
  Phys.\ Lett.\ B {\bf 713} (2012) 68
  [arXiv:1202.4083 [hep-ex]].

\end{thebibliography}
\end{document}